\documentclass[a4paper,twocolumn,preprintnumbers,citeautoscript,aps,prl,longbibliography,superscriptaddress,floatfix,nofootinbib]{revtex4-2}
\usepackage{amsmath,amsfonts,amssymb,graphics}
\usepackage[normalem]{ulem}
\usepackage{url}
\usepackage{graphicx}

\usepackage{mathrsfs}
\usepackage{bbm}
\usepackage{pifont}
\usepackage{braket}
\usepackage{siunitx}
\usepackage{slashed}

\usepackage{mathtools}
\usepackage{xspace}
\usepackage{booktabs}
\usepackage{csquotes}
\usepackage[dvipsnames,svgnames]{xcolor}
\usepackage{tikz}
\usetikzlibrary{positioning,shapes,arrows}
\usetikzlibrary{decorations.pathmorphing}
\usetikzlibrary{decorations.markings}
\usepackage{standalone}

\usepackage{url}
\usepackage[bookmarks=false, breaklinks]{hyperref} 

\definecolor{nicered}{rgb}{0.5,.0,.0}
\definecolor{darkblue}{rgb}{0,.1,.9}
\definecolor{lightblue}{rgb}{0,.1,.6}
\definecolor{applegreen}{rgb}{0.55, 0.71, 0.0}
\definecolor{darkgreen}{rgb}{0.0, 0.2, 0.13}
 
\hypersetup{colorlinks, citecolor=darkgreen ,linkcolor=nicered, urlcolor= lightblue}

\begin{document}
\title{About electroweak domain walls in Majoron models}
\author{Maximilian Berbig}
\email{berbig@ific.uv.es}
\affiliation{Departament de Física Teòrica, Universitat de València, 46100 Burjassot, Spain}
\affiliation{Instituto de Física Corpuscular (CSIC-Universitat de València), Parc Científic UV, C/Catedrático José Beltrán, 2, E-46980 Paterna, Spain}

\date{\today}

\begin{abstract}
 Some time ago it was claimed in \enquote{Spontaneous Breaking of Lepton Number and Cosmological Domain Wall Problem} (Phys. Rev. Lett. 122, 151301 (2019)) that non-perturbative instantons of the weak interaction $\text{SU}(2)_\text{W}$ lead to the formation of domain walls in Majoron models owing to the anomaly of the spontaneously broken global lepton number $L$ symmetry $\text{U}(1)_L$ with respect to $\text{SU}(2)_\text{W}$. We point out that it has long been known, that this effect can be completely rotated away unless there is a source of explicit $B+L$ breaking present, where $B$ denotes baryon number. We further estimate the tiny instanton induced Majoron mass from $B+L$ breaking  and analyze the cosmological impact of such domain walls. In general this scenario does not lead to a cosmological catastrophe and we demonstrate that the tiny instanton induced mass can act as a bias term to collapse walls induced by a larger source of lepton number breaking. Alternatively this electroweak Majoron could act as dynamical dark energy.
\end{abstract}

\maketitle

\textbf{Introduction.\textemdash}
Domain Walls from the spontaneous breaking of a discrete symmetry \cite{Zeldovich:1974uw} are perhaps the most well-known topological defects (consult e.g. Ref.~\cite{Vilenkin:2000jqa} for an overview over this vast subject) and have haunted particle cosmologists for more than fifty years.
Numerical simulations in Refs.~\cite{Press:1989yh,Garagounis:2002kt,Leite:2012vn,Hiramatsu:2013qaa} indicate that domain walls reach a scaling regime, where their size reaches the scale of the cosmological horizon $1/H$  in terms of the Hubble rate $H$ and there would be an $\mathcal{O}(1)$ number of walls inside our Hubble volume. 
As the energy density in domain walls redshifts in a radiation dominated background as $1/R^2$, where $R$ is the cosmic scale factor, domain walls are expected to eventually dominate over both radiation ($1/R^4$) and matter ($1/R^3$). A domain wall dominated universe would evolve 
with a scale factor that grows with cosmic time as $t^2$ \cite{Zeldovich:1974uw} and such power-law inflation would affect primordial nucleosynthesis, the decoupling of cosmic microwave photons or the formation of large scale structure \cite{Gelmini:1988sf}. Moreover the presence of even a single horizon sized domain wall could potentially ruin the cosmological principles of homogeneity and isotropy, which hold to a very high degree as is evident from the observed anisotropic temperature fluctuations $\Delta T/T \simeq 10^{-5}$ of the cosmic microwave background (CMB) \cite{Planck:2018jri}. 

In order to avoid such cosmological disasters one typically has to either avoid the use of discrete symmetries altogether or use inflation to dilute the defects \cite{Guth:1980zm}.  Recent simulations indicate that a statistical population bias for the distribution of field values from inflationary fluctuations does not lead to the collapse of the domain walls due to super-horizon correlations \cite{Gonzalez:2022mcx}. Low temperature symmetry restoration \cite{Weinberg:1974hy,Vilenkin:1981zs,Ramazanov:2021eya,Babichev:2021uvl} is another remedy.
One can also render the discrete symmetry to be only approximate via the introduction of explicit breaking terms colloquially known as \enquote{bias terms}, which allow for the destruction of the walls \cite{Kibble:1976sj,Vilenkin:1981zs,Sikivie:1982qv}.
Recently such collapsing domain walls have become en vogue again, since they could offer an explanation (see e.g. \cite{Ferreira:2022zzo}) for the correlated stochastic gravitational wave signal detected by various pulsar timing collaborations \cite{NANOGrav:2023gor,EPTA:2023fyk,Reardon:2023gzh,Xu:2023wog,InternationalPulsarTimingArray:2023mzf}.

One class of models that are affected by the domain wall problems are global $\text{U}(1)$ symmetries that are explicitly broken to a $\mathcal{Z}_N, \; N\in \mathbb{N}$. The most prominent offender is the QCD axion \cite{PhysRevLett.40.223,PhysRevLett.40.279} originating as the pseudo-Nambu-Goldstone boson (pNGB) of a  global Peccei-Quinn (PQ) symmetry  \cite{PhysRevLett.38.1440,Peccei:1977ur} that is anomalous and hence not a symmetry of the full quantum theory. Non-perturbative QCD instanton effects generate a potential for the QCD axion around the time of the QCD phase transition and domain walls appear once the axion has relaxed to one of the $N$ minima of its potential. On top of that cosmic strings are produced at the much earlier PQ phase transition \cite{Kibble_1976,PhysRevD.26.435,KIBBLE1980183}  and become the boundaries of the domain walls. For $N=1$ it was shown that the resulting hybrid defect will collapse on its own as the strings can cut the walls into smaller slices \cite{Vilenkin:1982ks,Barr:1986hs}  or via the nucleation of expanding closed string loops \cite{PhysRevD.26.435}.
The bias term solution can work here as well \cite{Sikivie:1982qv}, but faces pressure from the axion quality problem \cite{Beyer:2022ywc}. Alternatively a second interim contribution to the axion potential, that switches of before the QCD phase transition, can lead to a uniform axion field value in all Hubble patches \cite{Barr:2014vva,Reig:2019vqh,Zhang:2023gfu}. 

Global lepton number $\text{U}(1)_L$, that can explain the smallness of Majorana neutrino masses via the fermion-mediated Type I/III seesaw mechanisms \cite{Minkowski:1977sc,Yanagida:1979as,Gell-Mann:1979vob,Glashow:1979nm,Yanagida:1980xy, PhysRevLett.44.912, Foot:1988aq}  or the electroweak triplet scalar-mediated  Type II seesaw \cite{Lazarides:1980nt,Schechter:1980gr,Mohapatra:1980yp,PhysRevD.22.2860,Wetterich:1981bx},  is another popular choice of symmetry with its own associated pNGB called the Majoron \cite{Chikashige:1980qk,Chikashige:1980ui,Gelmini:1980re}. While the model in which the Majoron originates from the scalar triplet \cite{Gelmini:1980re} is excluded by the observed decay width of the $Z$-boson at LEP 1 \cite{Berezhiani:1992cd}, one can construct viable models via the inclusion of an additional scalar singlet \cite{Schechter:1981cv,Choi:1989hj,Choi:1991aa} that predominantly houses the Majoron. An overview over the Majoron phenomenology and its potential as a dark matter candidate can be found in Refs.~\cite{Garcia-Cely:2017oco,Brune:2018sab,Heeck:2019guh,Reig:2019sok,Akita:2023qiz}; the connection to proton decay was discussed in Ref.~\cite{Greljo:2025suh}.

Some time ago Ref.~\cite{Lazarides:2018aev} claimed that due to the well known ABJ anomaly \cite{Adler:1969gk,Bell:1969ts} of lepton number with respect to the weak gauge interaction  $\text{U}(1)_L\otimes \text{SU}(2)_\text{W}^2$  domain walls should be formed in Majoron models via weak interaction instanton processes  that typically  break lepton number down to a $\mathcal{Z}_3$. The authors of Ref.~\cite{Lazarides:2018aev} then focused on building Majoron models that either feature $N=1$ for a collapsing string-wall network or embed the remaining $\mathcal{Z}_N$ into a gauge-group \cite{Lazarides:1982tw,Barr:1982bb}, which makes all $N$ minima physically equivalent and thus prevents the formation of domain walls in the first place. This work then stimulated further research into building  Majoron models with $N=1$ and their cosmic implications \cite{Brune:2022vzd,Brune:2025zwg}. 

In this manuscript we demonstrate that the conclusions of Ref.~\cite{Lazarides:2018aev} are flawed because the authors overlooked the fact that electroweak instantons do not generate a potential for the Majoron in the absence of \textit{additional} explicit breaking of \textit{both} $B$ and $L$ in the linear combination $B+L$, where $B$ denotes baryon number. Furthermore, even if such a breaking was present we show that the correct Majoron potential from electroweak instantons depends on an \textit{additional} coupling constant and does not necessarily lead to   cosmological problems, even when considering finite temperature effects. The symmetry argument we provide has already been made in various forms throughout the years
\cite{Anselm:1992yz,Anselm:1993uj,Dvali:2005an,FileviezPerez:2014xju,Long:2015vsa,Shifman:2017lkj,Heeck:2019guh}
(see also Refs.~\cite{Quevillon:2019zrd,Quevillon:2020hmx,Quevillon:2020aij,Csaki:2023ziz} for examples from the QCD axion literature and Refs.~\cite{Nomura:2000yk,McLerran:2012mm,Ibe:2018ffn} from the dark energy literature).\footnote{In our analysis we do not take gravity into account; for more information consult Refs.~\cite{Dvali:2005an,Dvali:2024zpc}.}
While this line of reasoning is surely well known to experts, we think that a pedagogical summary will be useful to the particle physics and cosmology communities at large. 
In the next section we argue why the Majoron does not receive a potential from electroweak instantons  as long as $B+L$ is not explicitly broken and further  we estimate the Majoron mass in the presence of $B+L$ breaking from a dimension six operator. We then investigate the cosmological consequences of the Majoron mass and the associated domain walls. If there exists another less suppressed source of explicit lepton number breaking, that gives rise to a larger Majoron mass, the comparatively tiny contribution from the electroweak instanton can be used as a bias term to make the domain wall network collapse. In case the electroweak instanton is the leading contribution to the Majoron potential, then the Majoron can play the role of dynamical dark energy. We delineate the parameter space for both scenarios.

\begin{figure}
    \centering
    \includegraphics[width=0.7\linewidth]{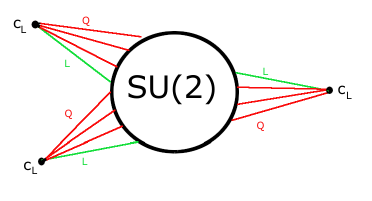}
    \caption{Diagrammatic representation of the single instanton 't~Hooft vertex with the fermion zero modes for each generation of $L\;(Q)$ in green (red) closed up by three insertions of the $B+L$ breaking effective operator in Eq.~\eqref{eq:B+L} with coefficient $c_L$. Adapted from Ref.~\cite{Csaki:2023ziz} (see also the related discussion in the appendix).}
    \label{fig:vertex}
\end{figure}
 
\textbf{The Argument.\textemdash}
The topological term of the weak interaction is given by 
\begin{align}
    -\frac{g_W^2 \theta_\text{EW}}{32\pi^2} W_{\mu\nu} \tilde{W}^{\mu\nu},
\end{align}
where $g_W$ is the gauge coupling and $\theta_\text{EW}$ is an angle. We can rephase  the lepton doublet $L$ (quark doublet $Q$) by a lepton number $\text{U}(1)_L$ (baryon number $\text{U}(1)_B$) transformation $L\rightarrow e^{-i \alpha_L} L\quad \left(Q\rightarrow e^{-i \alpha_B/3} Q\right)$ which induces a shift in the electroweak vacuum angle via the  $\text{U}(1)_{B+L}\otimes \text{SU}(2)_\text{W}^2$ anomaly \cite{FileviezPerez:2014xju} 
\begin{align}
    \theta_\text{EW} \rightarrow \theta_\text{EW} + N_g\left(\alpha_L +\alpha_B\right),
\end{align}
where $N_g=3$ is the number of Standard Model (SM) fermion generations.
The electroweak topological term can be removed for $\alpha_L+\alpha_B = -\theta_\text{EW}/N_g$. Now suppose we break lepton number spontaneously via the vev of a complex scalar $\sigma$ with a $\text{U}(1)_L$ charge of minus two  in  a single generation toy model based on e.g. the Type I Seesaw framework\footnote{The choice of high-scale Seesaw and number of generations is just for illustration and none of our conclusions depend on it.} with a sterile neutrino $N$\cite{Chikashige:1980qk,Chikashige:1980ui}\footnote{We utilize two component spinor notation \cite{Dreiner:2008tw} throughout this work.}
\begin{align}
    Y_N \sigma N N + Y_L L H N + \text{h.c.},
\end{align}
where we decompose $\sigma$ in terms of its radial (angular) mode $h_\sigma\;(j)$ and vacuum expectation value (vev) $v_L$
\begin{align}\label{eq:onehalf}
    \sigma = \frac{h_\sigma + v_L}{\sqrt{2}} e^{i\frac{j}{v_L}}.
\end{align}
Without loss of generality we take $Y_N$ to be real and carry out an anomaly-free redefinition of the gauge singlet  neutrinos $N$ as $N\rightarrow N e^{- i 
\alpha_N}$ with $\alpha_N =j/(2 v_L)$, which moves the Majoron into the $Y_L$ Yukawa interaction, and generates derivative couplings of $j$ to $N$. A field redefinition of $L$ with $\alpha_L = \text{Arg}[Y_L]-j/(2 v_L)$ moves the Majoron and the phase of the coupling $Y_L$ into the topological term, and generates derivative couplings of $j$ to $L$. Since Baryon number $B$ is unbroken we have the freedom to completely remove this term via a redefinition of $Q$, which also picks up a derivative coupling to $j$, with $\alpha_B =-(\theta_\text{EW}+\text{Arg}[Y_L])  + j/(2 v_L)$. Thus the Majoron   coupling to $W\tilde{W}$ is unphysical  and it can not become massive via $\text{SU}(2)_\text{W}$ instantons. The situation changes once we switch on  a term that also breaks baryon number and specifically $B+L$ like e.g. (see Refs.~\cite{Quevillon:2020hmx,Csaki:2023ziz,Beneito:2023xbk,Gargalionis:2024nij} and references therein for more operators)
\begin{align}\label{eq:B+L}
    -\mathcal{L}_{B+L} = \frac{c_L^{ijkl}}{M_\text{UV}^2} L_i Q_j Q_k Q_l,
\end{align}
which can arise in e.g. $\text{SU}(5)$ GUT \cite{Georgi:1974sy} from the exchange of gauge bosons or colored Higgses, where $c_L$ would be given by a gauge or Yukawa coupling squared and the cut-off scale $M_\text{UV}$ would be identified with the mass of the exchanged boson.
In the following we assume that $\text{SU}(2)_\text{W}$ is not unified with $\text{SU}(3)_c$ in the  $\text{SU}(5)$ GUT, because the $\text{U}(1)_L \otimes \text{SU}(2)_\text{W}^2$
anomaly would then also imply an anomaly with QCD (see also the discussion in Ref.~\cite{Shifman:2017lkj}). This would render the Majoron to be a QCD axion with its own distinct phenomenology \cite{Langacker:1986rj,Shin:1987xc,Ballesteros:2016euj,Ballesteros:2016xej}. Nevertheless unification can still be realized if $\text{U}(1)_L$ is an emergent accidental symmetry that only appears below the unification scale. Exotic processes like proton decay can be avoided by an appropriate flavor structure in the couplings of Eq.~\eqref{eq:B+L}, but the details will not be relevant for our discussion. A subtlety related to the operator in Eq.~\eqref{eq:B+L} will be elaborated upon in the appendix.

Redefining $Q$ with the aforementioned value for $\alpha_B$ just moves the Majoron from the topological term into $\mathcal{L}_{B+L}$ and hence the Majoron dependence can not be redefined away. For this to occur \textit{both} baryon and lepton number need to be violated and the relevant linear combination $B+L$ is picked out by the anomaly structure of $\text{SU}(2)_\text{W}$ \cite{FileviezPerez:2014xju}. In this case the coupling of the Majoron to $W\tilde{W}$ is physical and it will be affected by $\text{SU}(2)_\text{W}$ instantons. This is entirely analogous to the case of a QCD axion $a$ with massive quarks $q$, where chiral field redefinitions just move the axion back and forth between the topological term $a/f_a G\tilde{G}$ and the chiral symmetry breaking mass term $m_q q_L q_R$. As usual only \textit{relative} phases are physical and for the Majoron $\mathcal{L}_{B+L}$ plays the role of $m_q q_L q_R$ \cite{FileviezPerez:2014xju}. It is well known that the QCD axion mass vanishes in the presence of at least one massless quark. The underlying reason is that in this case the axion is aligned with  an anomaly-free linear combination of $\text{U}(1)_\text{PQ}$ and the restored chiral symmetry of the massless quark $\text{U}(1)_q$.
The analogous observation for the Majoron is that in the absence of $B+L$ breaking there is an anomaly free linear combination
of $\text{U}(1)_{L}$ and $\text{U}(1)_{B+L}$ that the Majoron is aligned with, and this linear combination is nothing more than  $\text{U}(1)_{B-L}$. This also removes any potential confusion about whether one should impose a global $\text{U}(1)_L$ or $\text{U}(1)_{B-L}$ when building Majoron models. Therefore the Majoron potential must depend on a combination of  $B+L$ violating couplings such as $c_L$ in Eq.~\eqref{eq:B+L} (see also the discussion in the appendix).

We first discuss the zero temperature mass of the Majoron and later focus on finite temperature effects. 
Using the methods of Ref.~\cite{Csaki:2023ziz} for estimating the effect a of 't Hooft determinant in a one-instanton-background  \cite{tHooft:1976rip,tHooft:1976snw} we can estimate the Majoron mass induced by the non-perturbative $\text{SU}(2)_\text{W}$ effects in the presence of explicit $B+L$ breaking. Throughout this paper we do not consider other sources of explicit symmetry breaking for lepton number. The limits for the integral over the instanton size $\rho$ are the  ultraviolet (UV) cutoff $1/M_\text{UV}$ of our effective theory as well as the infrared (IR) cutoff $1/(g_W v_H)$, where $v_H=\SI{246}{\giga\electronvolt}$ denotes the Higgs vev. The IR cutoff can be understood from the formalism of \enquote{constrained instantons} \cite{Affleck:1980mp} since due to the spontaneous breaking of $\text{SU}(2)_\text{W}$ there appears a suppression factor of $e^{-8 \pi^2 \rho^2 v_H^2}$, which effectively suppresses instantons larger than $1/(g_W v_H)$. Ref.~\cite{Csaki:2023ziz} obtained that the integral  for a $\text{SU}(2)$ gauge theory is always dominated by the UV cutoff, and these UV instantons are also known as \enquote{small size} instantons. This is different from the case of the QCD axion, whose potential is dominated by IR effects and not expected to be captured by a one instanton calculation as it is connected to confinement (see Ref.~\cite{Csaki:2023ziz} and references therein). 
Assuming the same renormalization running as in the Standard Model with the one-loop beta function $\beta_W=-b_0 g_W^3/(16\pi^2)$ and $b_0^\text{SM} = 19/6$ the estimate for the Majoron mass at zero temperature turns out to be \cite{Csaki:2023ziz}\footnote{Here we rewrote Eqn.~(4.7) of Ref.~\cite{Csaki:2023ziz} by using the definition of the RGE invariant scale in Eq.~(2.7) of the same reference, which explains the appearance of the exponential factor and the power of $M_\text{UV}$.}
\begin{widetext}
\begin{align}\label{eq:EW}
    m_{j}^2 \simeq \left(\SI{5e-29}{\electronvolt}\right)^2   |c_L|^3 \cos{(\theta_\text{EW}+ 3 \delta_L )} 
    \left(\frac{g_W\left(\SI{e16}{\giga\electronvolt}\right)}{g_W(M_\text{UV})}\right)^8
    \left(\frac{e^{-\frac{8\pi^2}{g_W^{2}(M_\text{UV})}}}{10^{-137}}\right)
    \left(\frac{\SI{e8}{\giga\electronvolt}}{v_L}\right)^2 \left(\frac{M_\text{UV}}{\SI{e16}{\giga\electronvolt}}\right)^4,
\end{align}
\end{widetext}
where we used that $g_W\left(\SI{e16}{\giga\electronvolt}\right)\simeq 0.523$  and the smallness can be understood from the exponential suppression of the instanton $\sim \exp{(-8\pi^2 /g_W\left(M_\text{UV}\right)^2)}$. It is evident from figure \ref{fig:vertex} that the Majoron mass requires three insertions of the operator in Eq.~\eqref{eq:B+L}, where for simplicity's sake we assumed flavor-blind couplings and defined $\delta_L \equiv \text{Arg}(c_L)$.  Including a viable flavor structure to  e.g. suppress proton decay will lead to additional suppression factors and might introduce a dependence on quark or lepton mixing angles as noted in Ref.~\cite{Nomura:2000yk}, but a detailed analysis of these implications is beyond the scope of this work.
If one uses operators built purely from $\text{SU}(2)_\text{W}$ singlet fields instead of Eq.~\eqref{eq:B+L}, the estimate for $m_{j}$ will be further reduced by  products involving all SM Yukawa couplings \cite{Csaki:2023ziz}.  
Note that due to the UV sensitivity this result is expected to change when $\text{SU}(2)_\text{W}$ is embedded in a larger group or in the presence of additional charged matter. The impact of adding additional scalar multiplets will be discussed around Eq.~\eqref{eq:increase}. The putative UV completion of the effective operator in Eq.~\eqref{eq:B+L} could also change this estimate.
Here we chose $v_L > \SI{e8}{\giga\electronvolt}$ to comply with astrophysical bounds from stellar cooling \cite{Heeck:2019guh} and neutrino masses from e.g. a Type I/III Seesaw can be realized with natural values for the Yukawa couplings $Y_L$ for $v_\sigma $ as large as $\SI{e15}{\giga\electronvolt}$.

\textbf{Cosmology.\textemdash}
First we show that the instanton induced mass in Eq.~\eqref{eq:EW} by itself only leads to domain walls that are not in conflict with standard cosmology.
For $m_j<\mathcal{O}(\SI{e-28}{\electronvolt})$ coherent oscillations of the Majoron \cite{Preskill:1982cy,Abbott:1982af,Dine:1982ah} occur after recombination so $j$ can not be the dark matter. We also checked that the corresponding relic density would be more than subdominant. The Majoron begins to oscillate when $H\simeq 3 m_j$, which implies $T\simeq \sqrt{m_j M_\text{Pl.}}$ during radiation domination. Once the Majoron relaxes to the minimum of its effective potential the $\mathcal{Z}_{N_g}=\mathcal{Z}_3$, that is preserved by the weak instantons (see Ref.~\cite{Koren:2022bam} for an overview over the anomaly free discrete symmetry of the SM), is spontaneously broken and domain walls appear as defects in the Majoron field. 
We estimate the surface tension  $\sigma_\text{DW}$ of a single Majoron domain wall  as \cite{Hiramatsu:2010yn,GrillidiCortona:2015jxo}
\begin{align}\label{eq:surf}
    \sigma_\text{DW} \simeq 9 m_j v_L^2,
\end{align}
and take the energy density of this horizon sized defect to be $\rho_\text{DW}\simeq \sigma_\text{DW} H$. 
The present day energy density fraction of a single domain wall is determined to be
\begin{align}
     \Omega_\text{DW} h^2
    \simeq  2\times10^{-16} h |\tilde{c}_L|^\frac{3}{2}  \left(\frac{v_L}{\SI{e8}{\giga\electronvolt}}\right) \left(\frac{M_\text{UV}}{\SI{e16}{\giga\electronvolt}}\right)^\frac{5}{12}\label{eq:omega},
\end{align}
where we defined $|\tilde{c}_L| =|c_L|\cos{(\theta_\text{EW}+ 3 \delta_L)}^{1/3}$, suppressed the dependence on $g_W$ purely for notational brevity and took $H_0 =h\times 100~\text{km}/(\text{s}\;\text{Mpc})$.
We checked numerically that the corresponding energy density is too small to ever dominate the energy budget of the universe in the past or today.
If the domain walls are present during BBN, their energy density can modify the rate of the background expansion similar to an additional species of dark radiation. Note that the equation of state parameter for a  domain wall ranges between $\omega=-2/3$ for a static wall \cite{Zeldovich:1974uw} and $\omega=1/3$ for a ultra-relativistic wall \cite{Gelmini:1988sf}, which means that its energy density can redshift  differently from radiation.
We recast limits on the abundance of  dark radiation  for a domain wall population  \cite{Hiramatsu:2010yn} and the  corresponding number of additional effective neutrinos reads
\begin{align}
    &\Delta N_\text{eff.}(T_\text{BBN}) 
    \simeq \nonumber\\&  7\times10^{-33} |\tilde{c}_L|^\frac{3}{2} 
    \left(\frac{v_L}{\SI{e8}{\giga\electronvolt}}\right) \left(\frac{M_\text{UV}}{\SI{e16}{\giga\electronvolt}}\right)^\frac{5}{12}\label{eq:neff},
\end{align}
where we set $T_\text{BBN}=\SI{1}{\mega\electronvolt}$. At the time of recombination with $T_\text{rec.}=\SI{0.1}{\electronvolt}$ we obtain $ \Delta N_\text{eff.}\simeq 4\times 10^{-19}$ for the same parameter choice. The presence of domain walls breaks the isotropy and homogeneity of the cosmological background. Such walls affect the propagation of photons via gravitational lensing, which can be expressed as a fluctuation of the cosmic background temperature $\Delta T/T\simeq \sigma_\text{DW} /(H M_\text{Pl.}^2)$ \cite{Zeldovich:1974uw, Friedland:2002qs} and one needs to check that this effect is not larger than the  observed CMB temperature anisotropies of $\Delta T/T \simeq 10^{-5}$. Numerically we obtain a present day value of 
\begin{align}
    \frac{\Delta T}{T}\simeq 9\times 10^{-17} |\tilde{c}_L|^\frac{3}{2}  \left(\frac{v_L}{\SI{e8}{\giga\electronvolt}}\right) \left(\frac{M_\text{UV}}{\SI{e16}{\giga\electronvolt}}\right)^\frac{5}{12}\label{eq:deltaT},
\end{align}
and it is obvious from Eqns.~\eqref{eq:omega}, \eqref{eq:neff} and \eqref{eq:deltaT} that the contribution of the Majoron domain wall to the energy budget of the universe at late times is completely negligible, unless $m_j$ is significantly enhanced by tens of  orders of magnitude from the aforementioned UV effects. 

In  Ref.~\cite{Lazarides:2018aev} it was claimed that thermal sphaleron transitions instead of the quantum tunneling encoded in the instanton  effect were responsible for the formation of the topological defect and it was assumed that $\sigma_\text{DW}\simeq m_W v_L^2$. This choice was justified by identifying the domain wall's width with the width of a sphaleron $\rho\sim 1/m_W$ \cite{Arnold:1987zg} (see also the discussion about the IR cutoff for a constrained instanton above Eq.~\eqref{eq:EW}). On the one hand this relation for the surface tension would imply $m_j\simeq m_W$, which is missing the required insertion of the $B+L$ violating couplings such as $c_L$, and numerically this guess is in stark disagreement with our estimate in Eq.~\eqref{eq:EW} by around forty orders of magnitude. On the other hand impact of QCD sphalerons on the axion dynamics was studied in Ref.~\cite{McLerran:1990de}, where it was found that the sphaleron effect does not alter the axion potential, but rather manifests itself as an additional source of friction in the equation of motion for this field. Hence for the Majoron we also do not expect sphalerons to play a role when it comes to its  potential and the domain wall formation.

\begin{figure*}[t]
    \centering
    \includegraphics[width=0.45\textwidth]{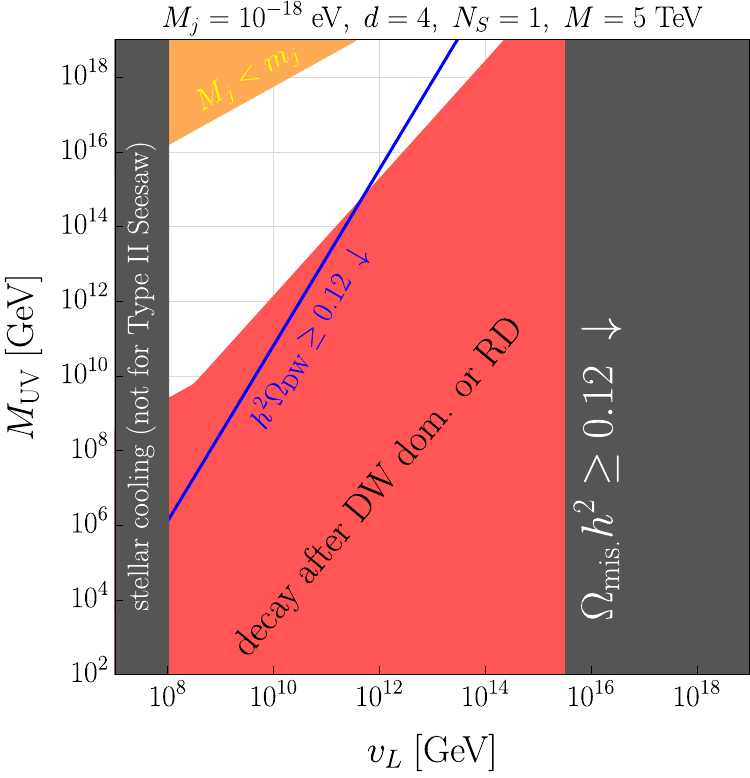}
    \hspace{0.05\textwidth}
    \includegraphics[width=0.45\textwidth]{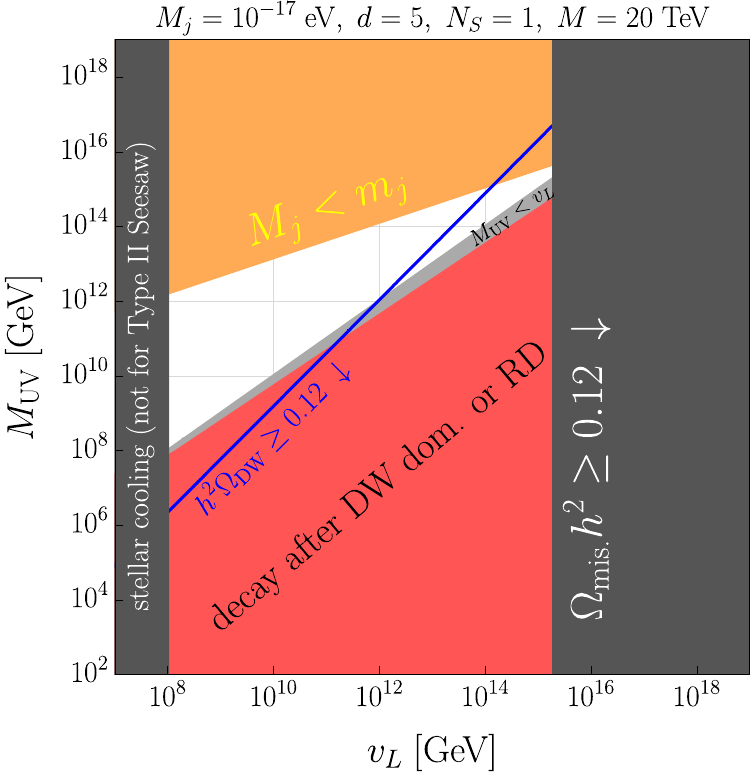}
    \caption{Parameter space for the production of ultra-light Majoron dark matter from domain wall decay. Here we depict a Majoron mass of $M_j=\SI{e-18}{\electronvolt}$ together with a single scalar quadruplet $(d=4)$ with mass $M=\SI{5}{\tera\electronvolt}$ contributing to the one-loop $\beta$-function of $\text{SU}(2)_\text{W}$ for the electroweak instanton induced bias term $m_j$ \textit{(left)} and $M_j=\SI{e-17}{\electronvolt}$ together with a quintuplet $(d=5)$ of mass $M=\SI{20}{\tera\electronvolt}$ \textit{(right)}. On the blue line one can reproduce the dark matter relic abundance solely by domain wall decay (see Eq.~\eqref{eq:OmegaBias}). The red region is ruled out because the domain walls would either decay after they dominate the universe's energy budget (see Eq.~\eqref{eq:cond1}) or after matter-radiation-equality (see Eq.~\eqref{eq:cond2}). In the orange slice the bias term contribution $m_j$ becomes larger than $M_j$. The small gray region is excluded because the lepton number breaking scale $v_L$ would need to be larger than the cut-off $M_\text{UV}$ of our EFT. Stellar cooling bounds from the Majoron's coupling to electrons typically rule out $v_L<\SI{e8}{\giga\electronvolt}$ for the Type I/III seesaw \cite{Heeck:2019guh}, but this can be avoided for the Type II seesaw \cite{Choi:1989hi}.  For larger values of $v_L>\SI{e15}{\giga\electronvolt}$ the energy density from coherent oscillations from the misalignment mechanism \cite{Preskill:1982cy,Abbott:1982af,Dine:1982ah} becomes the  dominant source of the dark matter relic density. In all plots we take $|c_L|\cos{(\theta_\text{EW}+ 3 \delta_L)}^{1/3}=1$.}
    \label{fig:ULDM}
\end{figure*}

\textbf{Bias term.\textemdash}
In case another source of explicit lepton number symmetry breaking (e.g. quantum gravity \cite{Georgi:1981pu,Dine:1986bg,Coleman:1989zu,Abbott:1989jw,Holman:1992us,Kamionkowski:1992mf,Barr:1992qq,Ghigna:1992iv,Kallosh:1995hi,Alonso:2017avz}) is responsible for   significantly larger Majoron masses and thus domain wall tensions, we could use the comparatively tiny contribution from the weak anomaly in Eq.~\eqref{eq:B+L} as a bias term to make these walls collapse \cite{Preskill:1991kd,Chiang:2020aui,Kitajima:2023cek,Bai:2023cqj,Blasi:2023sej,Lu:2023mcz}. This ansatz can be compared to case of the QCD axion, where the bias term arises from additional explicit breaking by hand or from quantum gravity, whereas the dominant contribution to the potential is due to QCD instantons. For the Majoron the roles are reversed since the bias term will come from electroweak instantons and the leading contribution to the potential originates  from explicit breaking by hand or quantum gravity. Moreover the strength of the bias term for the QCD axion is limited from above (see e.g. Ref.~\cite{Beyer:2022ywc}) by the shift of the potential minimum $\overline{\theta}$ away from zero, which should be $\overline{\theta}<10^{-10}$ to still solve the strong CP problem, whereas no such constraint exists for the Majoron as no known observable depends on the rephasing invariant electroweak vacuum angle $\theta_\text{EW}+ 3 \delta_L$.

First we denote the dimensionless Majoron field as $\theta_j\equiv j/(2 v_L)$ due to the residual $\mathcal{Z}_2$ symmetry of the seesaw models (see the redefinition of $N$ below Eq.~\eqref{eq:onehalf}). The electroweak potential is $\propto \cos{(3\theta_j)}$ due the $\text{U}(1)_{B+L}\otimes \text{SU}(2)_\text{W}^2$ anomaly for $N_g=3$ generations of SM fermions. We parameterize the additional and dominant source of lepton number breaking with an operator of the form 
\begin{align}\label{eq:V}
    V_n=  c_n \frac{\sigma^n}{M_\text{Pl.}^{n-4}} + \text{h.c.}
\end{align}
which gives rise to a potential $\propto \cos{(2 n \theta_j+\text{Arg}(c_n))}$ and we take the mass $M_j^2 \simeq 2 c_n v_L^2 (v_L/M_\text{Pl.})^{n-4} \sin{(\text{Arg}(c_n))}$ originating from this term to dominate over the instanton induced one $M_j\gg m_j$. In this scenario the domain wall surface tension now reads approximately $9 M_j v_L^2$.  As long as $n$ is not co-prime\footnote{That means the greatest common divisor of $n$ and 3 is 1.} with $3$  the electroweak contribution will bias the previously degenerate minima of $V_n$. As long as $m_j/M_j<0.2$ the different vacua will percolate sufficiently \cite{Gelmini:1988sf,Hiramatsu:2010yz,Hiramatsu:2010yn}. The expansion of the energetically lower and therefore true vacuum will push the domain walls towards each other, eventually leading to the collapse of the network. In order for this to occur before the walls dominate the energy density of the universe we have to demand that 
\begin{align}\label{eq:cond1}
    m_j > \SI{5.7e-23}{\electronvolt} \left(\frac{M_j}{\SI{e-18}{\electronvolt}}\right) \left(\frac{v_L}{\SI{e13}{\giga\electronvolt}}\right).
\end{align}
The energy density of the domain wall is then released in the form of an abundance of Majorons with \cite{Reig:2019sok}
\begin{align}\label{eq:OmegaBias}
    \Omega_\text{bias} h^2&\simeq \nonumber\\ 0.12 &\left(\frac{M_j}{\SI{e-18}{\electronvolt}}\right)^\frac{3}{2} \left(\frac{\SI{1.7e-21}{\electronvolt}}{m_j}\right)\left(\frac{v_L}{\SI{e13}{\giga\electronvolt}}\right)^2.
\end{align}
Here we focus on ultralight Majorons, that can be good dark matter candidates and their mass range is constrained by various astrophysical and cosmological considerations (see Ref.~\cite{Eberhardt:2025caq} for an overview). Structure formation bounds from the Lyman-$\alpha$ forest rule out $M_j< \SI{2e-20}{\electronvolt}$ \cite{Rogers:2020ltq}. Some more stringent limits come from angular momentum loss of rotating black holes via the superradiance effect. Observations of spinning supermassive black holes exclude masses in the window $\SI{7e-19}{\electronvolt}<M_j<\SI{e-16}{\electronvolt}$ \cite{Stott:2018opm}. However these limits should be interpreted carefully due to uncertainties in the measurements of the black hole spins \cite{Baryakhtar:2025jwh}. Furthermore the previous bounds assumed no self interactions, which constrains the decay constant for a conventional QCD axion to be above $\SI{e14}{\giga\electronvolt}$  \cite{Arvanitaki:2014wva}. Since the Majoron potential scales differently with its decay constant $\sim v_L$ compared to a QCD axion and we typically consider $v_L<\SI{e15}{\giga\electronvolt}$ for the Seesaw mechanism, we do not impose this bound and only consider the constraint from the Lyman-$\alpha$ forest. 
It is important to emphasize that this ultra-light mass range is difficult to realize for both a field-theoretic\footnote{The non-minimal model of Refs.~\cite{Hook:2018jle,DiLuzio:2021pxd} is a well known field-theoretic exception, that however suffers from other drawbacks \cite{Lu:2023ayc,Co:2025jnj}.} and a string-theoretic QCD axion \cite{Benabou:2025kgx}, which provides further motivation for our Majoron based setup.

In order for the Majorons produced in the domain wall collapse to constitute the dark matter, we should impose that they are produced before matter-radiation equality at the temperature $T\simeq \SI{0.76}{\electronvolt}$, which implies
\begin{align}\label{eq:cond2}
    m_j > \SI{5.9e-23}{\electronvolt}\sqrt{\frac{M_j}{\SI{e-18}{\electronvolt}}}.
\end{align}
Note that since the Majoron only couples very weakly to matter, there is no obstruction to having the domain walls decay around or after the time of BBN \cite{Reig:2019sok}. 
Equations \eqref{eq:cond1}-\eqref{eq:cond2} illustrate, that the instanton contribution in Eq.~\eqref{eq:EW} is too small for our purposes, if we only include which the contribution of the SM degrees of freedom to the $\text{SU}(2)_\text{L}$ $\beta$-function. This can be remedied by changing the running of weak gauge couplings, so that it decreases less in the UV, which softens the exponential suppression of $m_j$. Adding  $N_S$ additional scalar $\text{SU}(2)_\text{W}$ representations at the scale  $M<M_\text{UV}$ changes the one-loop $\beta$-function from $19/6$ to  $b_0^\text{BSM} < 19/6$ and thereby increases  \cite{Csaki:2023ziz}\footnote{Here we do not include additional fermions following Ref.~\cite{Csaki:2023ziz}, because they typically lead to additional zero modes in the one instanton background and thus can change the integral over the instanton radius. The insertions to close of the additional zero modes can even cancel the enhancement from the change of the running, which was found to occur in supersymmetric theories in Refs.~\cite{Nomura:2000yk,Ibe:2018ffn}.
Moreover the change in the prefactor $g_W$ of Eq.~\eqref{eq:EW} due to the additional scalars is a two loop effect \cite{Csaki:2023ziz}.}
\begin{align}\label{eq:increase}
    m_j^2 \rightarrow \left(\frac{M_\text{UV}}{M}\right)^{\frac{19}{6}-b_0^\text{BSM}} m_j^2.
\end{align}
We can increase the bias term for $M\ll M_\text{UV}$ and note that generically the new intermediate scale $M$ should lie above some TeV due to laboratory bounds on the new scalars.  
The change in the one-loop beta function for $N_S$ mass degenerate scalars in a $d$-dimensional $\text{SU}(2)_\text{W}$ representation reads \cite{Jurciukonis:2024cdy}
\begin{align}\label{eq:dim}
    \frac{19}{6}-b_0^\text{BSM} = N_S \frac{d(d^2-1)}{36},
\end{align}
which vanishes for a gauge singlet as expected. For $N_S=1$ tree level perturbative unitarity of $\text{SU}(2)_\text{W}$ gauge interactions requires $d<8\;(9)$ for a complex (real) scalar \cite{Hally:2012pu} and the $\text{SU}(2)_\text{W}$ gauge couplings remain perturbative below the Planck scale as long as $d<7$ \cite{Cirelli:2005uq}. Of course there exist other constraints on additional scalar representations such as electroweak precision observables or Higgs vacuum stability, but here we are merely interested in a proof of concept that demonstrates the existence of a suitable UV theory.

We depict the parameter space subject to the previous considerations  for two values of $M_j= \SI{e-18}{\electronvolt},\;\SI{e-17}{\electronvolt}$ in Fig.~\ref{fig:ULDM}. 
For both benchmarks we find that $\SI{e12}{\giga\electronvolt}<v_L<\SI{e15}{\giga\electronvolt}$ and that the cut-off $M_\text{UV}$ can remain safely below the Planck scale, possibly indicating a GUT origin for the effective operator in Eq.~\eqref{eq:B+L}. For $|c_L|\cos{(\theta_\text{EW}+ 3 \delta_L)}^{1/3}=1$ the aforementioned Majoron masses require e.g. a single generation of either a quadruplet $(d=4)$ with $M=\SI{5}{\tera\electronvolt}$ or a quintuplet $(d=5)$ with $M=\SI{20}{\tera\electronvolt}$ in order to have an early enough collapse and to explain the observed dark matter relic abundance. In case the bias term is further suppressed by $|c_L|\cos{(\theta_\text{EW}+ 3 \delta_L)}^{1/3}<1$, or due to the  flavor structure of $c_L$ we have to increase either $d$, $N_S$ or decrease $M$. For $v_L=\SI{e13}{\giga\electronvolt}$ one can realize $M_j=\SI{e-18}{\electronvolt}$ via an operator with dimension $n=17$ in Eq.~\eqref{eq:V}. As usual when dealing with the explicit breaking of a symmetry via ad-hoc higher dimensional operators we have to assume a mechanism to protect the required quality of the potential in Eq.~\eqref{eq:V} by  preventing the appearance of all lower dimensional operators with $n<17$ or suppressing their coefficients \cite{Georgi:1981pu,Dine:1986bg,Coleman:1989zu,Abbott:1989jw,Holman:1992us,Kamionkowski:1992mf,Barr:1992qq,Ghigna:1992iv,Kallosh:1995hi,Alonso:2017avz}.

One can deduce that the Majoron yield from the domain wall destruction in Eq.~\eqref{eq:OmegaBias} dominates over the misalignment contribution \cite{Preskill:1982cy,Abbott:1982af,Dine:1982ah} with a patch-averaged misalignment angle of $\theta_i\simeq 2.1$ \cite{GrillidiCortona:2015jxo} for $v_L<\SI{e15}{\giga\electronvolt}$. Since the energy density of gravitational waves radiated from the collapsing network scales with $\sigma_\text{DW}^2\sim M_j^2$ \cite{Hiramatsu:2012sc} we do not expect a detectable gravitational wave spectrum. Furthermore the collapse typically happens long after the QCD crossover, so the peak frequency of the spectrum would be too small to account for the signal observed in pulsar timing arrays \cite{NANOGrav:2023gor,EPTA:2023fyk,Reardon:2023gzh,Xu:2023wog,InternationalPulsarTimingArray:2023mzf}. 

\begin{figure*}[t]
    \centering
    \includegraphics[width=0.3\textwidth]{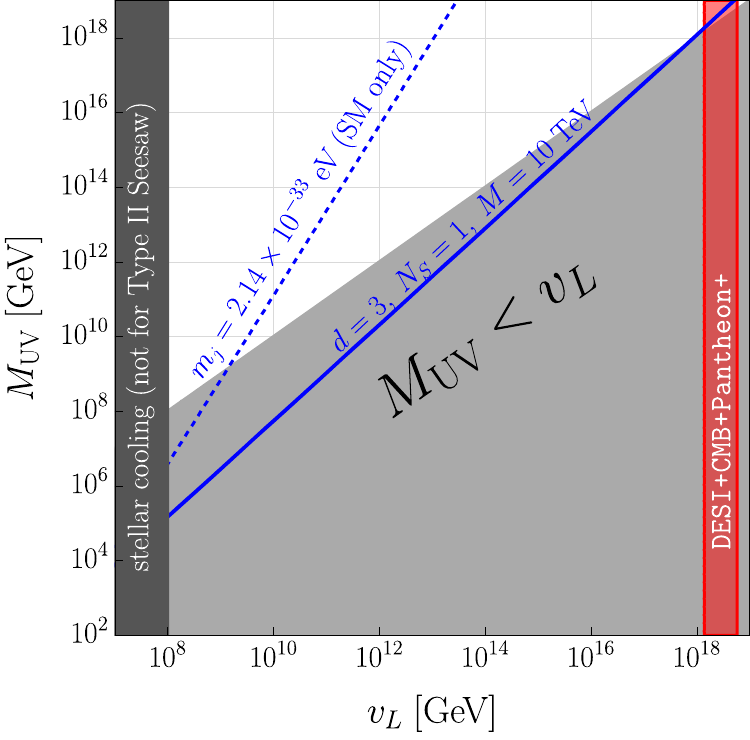}
    \hspace{0.03\textwidth}
    \includegraphics[width=0.3\textwidth]{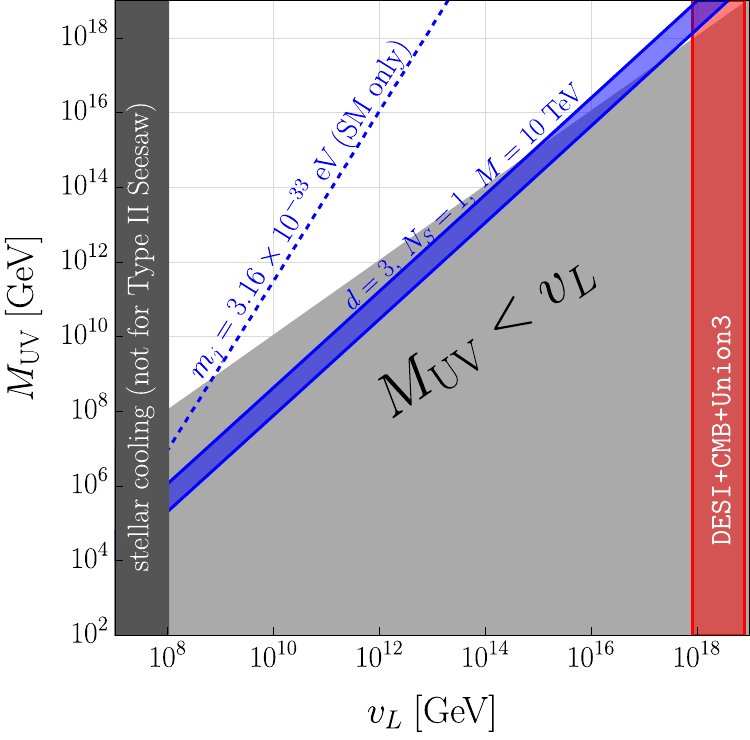}
    \hspace{0.03\textwidth}
     \includegraphics[width=0.3\textwidth]{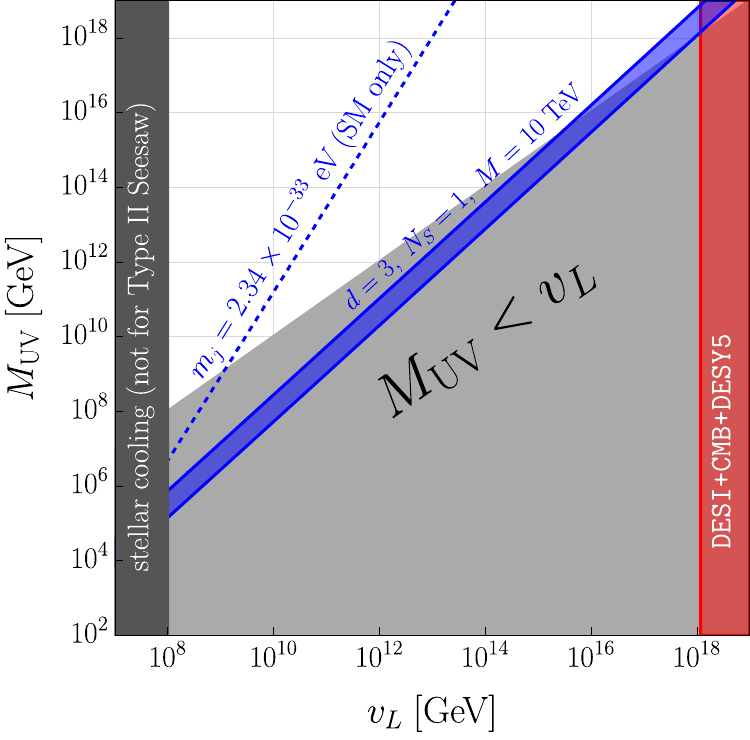}
    \caption{Parameter space that reproduces the preferred range of Majoron masses and decay constants $f_j\equiv 2 v_L/3$ for the hint of thawing quintessence in the combined data from \texttt{DESI} \cite{DESI:2024mwx,DESI:2024aqx,DESI:2024kob,DESI:2025zgx,DESI:2025fii}, the \texttt{CMB} surveys \texttt{Planck}  \cite{Planck:2018vyg,Planck:2019nip} and \texttt{ACT} \cite{ACT:2023dou,ACT:2023kun}, as well as the supernova catalogs \texttt{Pantheon+} \cite{Brout:2022vxf} \textit{(left)}, \texttt{Union3} \cite{Rubin:2023jdq} \textit{(middle)} and \texttt{DESY5} \cite{DES:2024jxu}  \textit{(right)} found by the analysis in Ref.~\cite{DESI:2025fii}. The blue bands correspond to the required rage of $m_j$ and the red bands indicate the needed range of $v_L$. In the gray region the lepton number breaking scale is larger than the UV cutoff $M_\text{UV}$ of the assumed effective field theory. In all plots we take $|c_L|\cos{(\theta_\text{EW}+ 3 \delta_L)}^{1/3}=1$ and include the effect of a single triplet $(d=3)$ scalar with a mass of $M=\SI{10}{\tera\electronvolt}$ on the one loop $\beta$-function of $\text{SU}(2)_\text{W}$.}
    \label{fig:DE}
\end{figure*}

\textbf{Dark Energy.\textemdash}
If the electroweak instanton induced potential is the sole or leading contribution to the Majoron mass, this scale is suggestively close the value of the Hubble rate today $H_0=\mathcal{O}(\SI{e-33}{\electronvolt})$, which implies that the Majoron could act as dynamical dark energy, also known as quintessence \cite{Fujii:1982ms,PhysRevD.35.2339,Wetterich:1987fm,Ratra:1987rm}, which was previously considered in Refs.~\cite{Fukugita:1994hq,Nomura:2000yk,McLerran:2012mm,Ibe:2018ffn}.
pNGB quintessence models \cite{Frieman:1995pm,Choi:1999xn,Nomura:2000yk,Kim:2002tq,Hall:2005xb,Barbieri:2005gj}  with https://www.overleaf.com/learn $m_j\simeq H_0$ typically require decay constants $f_j\equiv 2 v_L/3 \lesssim M_\text{Pl.}$ to explain the present day energy density about $0.7\times 3 H_0^2 M_\text{Pl.}^2/(8\pi)$ and such large $v_L$ would suppress the mass from Eq.~\eqref{eq:EW} far below $H_0$ for $M_\text{UV}$ at or below the Planck scale. This outcome can be avoided by again modifying the $\text{SU}(2)_\text{W}$ $\beta$-function via Eq.~\eqref{eq:dim}, which increases $m_j$ parametrically via the relation in Eq.~\eqref{eq:increase}.  Another way to come to the same conclusion is to note that the Majoron's potential for $M_\text{UV}\simeq M_\text{Pl.}$ reads $V_j= \mathcal{O}(10^{-132}M_\text{Pl.}^4)$ \cite{Ibe:2018ffn}, which is much smaller than the required energy density of the universe today of roughly $\mathcal{O}(10^{-123}M_\text{Pl.}^4)$ \cite{,McLerran:2012mm}.  Here the caveats from the quality problem \cite{Georgi:1981pu,Dine:1986bg,Coleman:1989zu,Abbott:1989jw,Holman:1992us,Kamionkowski:1992mf,Barr:1992qq,Ghigna:1992iv,Kallosh:1995hi,Alonso:2017avz} apply as well, and we assume some mechanism that guarantees that all other unwanted contributions to the Majoron potential are absent.

In fact recent  data from the first two releases of the \texttt{DESI} collaboration, which measures baryon acoustic oscillations, point towards a hint for time-dependent dynamical dark energy \cite{DESI:2024mwx,DESI:2024aqx,DESI:2024kob,DESI:2025zgx,DESI:2025fii}, whose  significance ranges between about $(3-4)\sigma$ depending on the supernova datasets used in the combined analysis. Attempts to explain this hint in terms of thawing \cite{Dutta:2008qn} pNGB quintessence were undertaken in Refs.~\cite{Tada:2024znt,Bhattacharya:2024kxp,Berbig:2024aee,DESI:2025fii,Urena-Lopez:2025rad,Lin:2025gne}. The analysis of Ref.~\cite{DESI:2025fii} found the following parameter ranges by combining the \texttt{DESI} data with CMB data from \texttt{Planck}  \cite{Planck:2018vyg,Planck:2019nip} and \texttt{ACT} \cite{ACT:2023dou,ACT:2023kun} with various supernova datasets:
$\log_{10}{(m_j/\SI{}{\electronvolt})}=-32.67^{+0.23}_{-0.25}$ and $\log_{10}{(f_j/M_\text{Pl.}^\text{red.})}=-0.13^{+0.33}_{-0.29}$ for \texttt{Pantheon+} \cite{Brout:2022vxf}, $\log_{10}{(m_j/\SI{}{\electronvolt})}=-32.50^{+0.28}_{-0.30}$ and $\log_{10}{(f_j/M_\text{Pl.}^\text{red.})}=-0.29^{+0.63}_{-0.35}$ for \texttt{Union3} \cite{Rubin:2023jdq}, as well as $\log_{10}{(m_j/\SI{}{\electronvolt})}=-32.63^{+0.26}_{-0.30}$ and $\log_{10}{(f_j/M_\text{Pl.}^\text{red.})}=-0.09^{+0.66}_{-0.40}$ for \texttt{DESY5} \cite{DES:2024jxu}. Here $M_\text{Pl.}^\text{red.}$ denotes the reduced Planck mass. 

From the parameter space depicted in figure \ref{fig:DE} we find that the hint for dynamical dark energy for each supernova dataset 
be accounted for by e.g.  the addition of a single triplet $(d=3)$ scalar with a mass of $M=\SI{10}{\tera\electronvolt}$ for $|c_L|\cos{(\theta_\text{EW}+ 3 \delta_L)}^{1/3}=1$. Note that the required value  $v_L=\mathcal{O}(\SI{e18}{\giga\electronvolt})$ is close to, or even above, the reduced Planck scale. String theory motivated conjectures such as the weak gravity conjecture \cite{Arkani-Hamed:2006emk} and the swampland distance conjecture \cite{Ooguri:2006in} seem to rule trans-Planckian decay constants. To be more precise the weak gravity conjecture demands for a given instanton action $S_\text{inst.}$ that \cite{Arkani-Hamed:2006emk}
\begin{align}
    f_j < \frac{M_\text{Pl.}^\text{red.}}{S_\text{inst}}.
\end{align}
Pushing the cutoff $M_\text{UV}$ to the reduced Planck scale  we find for the case of pure SM $S_\text{inst.}=8\pi^2/g_W^2(M_\text{Pl.}^\text{red.})=\mathcal{O}(100)$ and for the aforementioned modification of the running from a triplet with mass $\SI{10}{\tera\electronvolt}$ we numerically obtain an instanton action of the same order of magnitude. Hence the the lepton number breaking scale should be $v_L<\mathcal{O}(\SI{e16}{\giga\electronvolt})$ and we conclude that parameter space  for dark energy is incompatible with these conjectures. A supersymmetric solution of this problem can be found in Ref.~\cite{Ibe:2018ffn}.  Furthermore the preferred regions of parameter space in figure \ref{fig:DE} are located at the edge of the regime of validity $(v_L<M_\text{UV})$ for our effective treatment in terms of the operator in Eq.~\eqref{eq:B+L}. On the other hand this could also be interpreted as pointing towards Planck scale origin of Eq.~\eqref{eq:B+L}, via explicit quantum gravitational breaking of $\text{U}(1)_\text{B+L}$.

The previous discussion involved a vanishing cosmological constant and a quintessence field, whose equation of state parameter $\omega_j$ never crosses over into the so called  \enquote{phantom} regime $\omega_j<-1$ \cite{Caldwell:1999ew,Caldwell:2003vq}. While some authors (see e.g. Ref.~\cite{Gialamas:2024lyw}) claim that the apparent preference for a phantom crossing found in the first \texttt{DESI} data release could be attributed to the phenomenological CPL parameterization \cite{Chevallier:2000qy,Linder:2002et} of the equation of state used in many fits, non-parametric reconstruction methods also seem find a phantom behavior \cite{DESI:2025fii}. One can accommodate this feature via the assumption of a negative kinetic term for the quintessence field, which however leads to a violation of the null energy condition and various problems such as e.g. vacuum decay \cite{Cline:2003gs} (see Ref.~\cite{Ludwick:2017tox} for a review). Modified gravity theories that equip the quintessence field with a non-minimal coupling to gravity \cite{Ye:2024ywg,Wolf:2024stt,Wolf:2025jed} can describe the phantom crossing without running into the aforementioned pathologies. Another approach is to consider multiple fields, where each component has an equation of state above $-1$, but the \textit{effective} equation of state for the multi-field system can cross the phantom divide. Typical examples consider a coupling between quintessence and dark matter \cite{Amendola:1999er,Das:2005yj}, see Refs.~\cite{w4qb-plk8,Bedroya:2025fwh} for  recent microphysical implementations of this idea. In Ref.~\cite{Liu:2025bss} a very minimal realization of this \enquote{phantom mirage} was presented: Here one considers a cosmological constant $\Lambda$ together with a  Majoron of the mass  $\log_{10}{(m_j/\SI{}{\electronvolt})}=-32.5$ with a present day energy fraction of $r_j=\rho_j/(\rho_j+\rho_\Lambda)\simeq (8-12)\%$. Initially the Majoron will be frozen  and it acts  as a component of dark energy $(\omega_j\gtrsim -1)$ at redshifts $z\geq 1$. Due to  $m_j\gtrsim H_0$ it begin its coherent oscillations before the present time $z=0$, so that it acts as dark matter today $(\omega_j=0)$. The interplay of Majoron's dynamics with the  cosmological constant realizes the phantom crossing of their effective equation of state. Due to the smaller energy fraction $r_j\simeq (8-12)\%$ one requires a somewhat smaller lepton number breaking scale of $v_L\simeq (4.7-5.8)\times10^{17}\;\text{GeV}$ compared to the quintessence scenario without a cosmological constant. This makes it easier to accommodate the mirage scenario in our effective field theory approach, but we note that this scale is still in violation of the weak gravity conjecture \cite{Arkani-Hamed:2006emk} by about an order of magnitude. 

If the dark energy is realized in terms of our electroweak Majoron, we do not expect it to be able to explain the observed birefringence in the CMB \cite{Minami:2020odp,Diego-Palazuelos:2022dsq,Eskilt:2022cff,ACT:2025fju}, because $\text{U}(1)_\text{B-L}$ is anomaly free with respect to electromagnetism so that the Majoron-photon-coupling will be suppressed by the smallness of the Majoron mass \cite{Bauer:2017ris,Heeck:2019guh}.

\textbf{Conclusions.\textemdash}
We critically reexamined previous claims  about the formation of domain walls in theories of spontaneously broken global lepton number due to explicit breaking from non-perturbative effects of the weak interaction \cite{Lazarides:2018aev}. Based on a host of previous arguments \cite{Anselm:1992yz,Anselm:1993uj,FileviezPerez:2014xju,Long:2015vsa,Heeck:2019guh} we reiterated that the Majoron can only couple to the topological term of the weak interaction if $B+L$ is explicitly broken by \textit{additional} interactions. Using the methods of Ref.~\cite{Csaki:2023ziz} we estimated the Majoron  mass, which heavily depends on the details of the UV theory.
Typically the energy density of the resulting Majoron potential is completely negligible in the early universe. Consequently even \textit{if} $B+L$ is explicitly broken, the domain walls in Majoron models do \textit{not} automatically lead to cosmological disasters and the model building complications suggested in Ref.~\cite{Lazarides:2018aev} are not necessarily required.

The ultra-light electroweak Majoron could be relevant for  dark energy in the late universe \cite{Nomura:2000yk,McLerran:2012mm,Ibe:2018ffn} and might explain the hint for dynamical dark energy in the recent \texttt{DESI} data \cite{DESI:2024mwx,DESI:2024aqx,DESI:2024kob,DESI:2025zgx,DESI:2025fii}.  Alternatively if there exists a larger source of explicit lepton number breaking apart from the weak instantons, we can utilize the instanton induced potential as a bias term to destabilize the wall network, which can produce the correct dark matter relic density in terms of ultra-light Majorons. Thus the weak anomaly of lepton number could actually be a \textit{virtue} instead of disaster.

\textbf{Acknowledgments.\textemdash}
We would like to thank Timo Brune, Salvador Centelles Chuliá, Mario Reig Lopez  and José W. F. Valle for fruitful discussions as well as Juan Herrero-García and Andreas Trautner for comments on the manuscript. Additionally we are grateful to the anonymous referee for reminding us of Ref.~\cite{McLerran:1990de} and for clarifying the results therein.  The author would  sincerely like to thank Antonio Herrero-Brocal for pointing out that upon field redefinition the $B-L$ conserving operator in $\mathcal{L}_\text{B+L}$ does not actually pick up a Majoron dependence in the minimal model without additional vector-like fermions (see also the related Ref.~\cite{Herrero-Brocal:2026nmc}), which lead to the appendix.  MB is supported by \enquote{Consolidación Investigadora Grant CNS2022-135592}, and funded also by \enquote{European Union NextGenerationEU/PRTR}, as well as the Generalitat Valenciana APOSTD/2025 Grant No. CIAPOS/2024/148.

\appendix
\section{Appendix}\label{sec:app}

The operator in the Lagrangian $\mathcal{L}_\text{B+L}$ in  Eq.~(5) of the main text does not actually pick up a coupling to the Majoron when performing the field redefinitions of only  $L$ and $Q$ with the transformation parameters $\alpha_L$ and $\alpha_B$ indicated between Eqns.~\eqref{eq:onehalf} and \eqref{eq:B+L} of the main text. One can show that under these redefinitions $\mathcal{L}_\text{B+L}\rightarrow e^{-i (\alpha_B+ \alpha_L)} \mathcal{L}_\text{B+L}$ in terms of $\alpha_B + \alpha_L= -\theta_\text{EW}$. The problem is that the operator in Eq.~\eqref{eq:B+L} accidentally conserves $B-L$, whereas the Majoron coupling arises from $\alpha_B-\alpha_L=j/v_L - \theta_\text{EQ} - 2 \text{Arg}[Y_L]$. This situation can be rectified either by invoking a different operator breaking both $B-L$ and $B+L$ or by changing the anomaly structure of the theory by allowing for chiral $B-L$ charges (compared to the vector-like charge assignment of $B-L$ for the SM fermions):
We can add e.g. a pair of fermion doublets $\Psi_{L,R}\sim(\textbf{1},\textbf{2},\mp 1/2)$, which are vector-like under the SM gauge symmetries, but chiral under Lepton number, and obtain a mass $M_\Psi\equiv Y_\Psi v_L/\sqrt{2}$ from the spontaneous breaking of lepton number via the operator 
\begin{align}
   Y_\Psi \sigma \Psi_L \Psi_R.
\end{align}

Before continuing with the impact on the Majoron, let us comment on the phenomenological viability of the additional doublet fermions: In order to avoid a relic abundance of stable $\Psi$ we can either suppress their thermal production or make them unstable: The first option requires $T_
\text{RH}<M_{\Psi}$ for sufficient Boltzmann suppression, where $T_\text{RH}$ is the post-inflationary reheating temperature in the instantaneous decay approximation. Since the Majoron domain wall discussion in the main text assumes post-inflationary $B-L$ breaking we have to impose $v_L<T_\text{RH}< M_{\Psi}$, which amounts to a narrow range for the  Yukawa coupling close to its perturbative maximum $1<Y_\Psi/\sqrt{2}< \sqrt{2\pi}\simeq 2.5$. On the other hand we can also add a decay mode for $\Psi$. For instance with the normalization $Q_{B-L}[\sigma]=-2$ we can choose the charge  assignment  and $Q_{B-L}[\Psi_L]=-1,\; Q_{B-L}[\Psi_R] = 3$, which allows for the operators $\Psi_L H N$ and $\Psi_L \tilde{H} e$, where $e$ is a charged lepton singlet and $\tilde{H}$ the $\text{SU}(2)_\text{W}$ conjugate of $H$. Here all gauge and generation indices are suppressed. The first operator is only present in the Type I and Type III Seesaw models and the second operator works for all kinds of Seesaw models. If the Yukawa coupling of this operator is not suppressed one can have fast decays $\Psi\rightarrow H e^\dagger$, so we do not expect any $\Psi$ relics to be present at late times.

We can rotate the Majoron into the topological term of $\text{SU}(2)_
\text{L}$ via the transformations $\Psi_{L,R} \rightarrow e^{-i \alpha_{\Psi_{L,R}}} \Psi_{L,R}$ with $\alpha_{\Psi_L}+\alpha_{\Psi_R} = j/v_L$ which induces the shift of the topological angle of $\theta_\text{EW}\rightarrow \theta_\text{EW} + j/v_L$. The rotation of the lepton doublet in terms of $\alpha_L$ is left as in the main text, but now we can remove the Majoron from the topological term via a transformation of $Q$ with $\alpha_B=-\theta_\text{EW}-\text{Arg}[Y_L]-j/(2 v_L)$. This leads to $\alpha_B + \alpha_L= -\theta_\text{EW} - j/v_L$ so the Majoron now appears in the operator in Eq.~\eqref{eq:B+L} of the main text. In the presence of the vector-like pair of doublets additional fermionic zero modes arise that contribute to the 't Hooft vertex in Figure~\ref{fig:vertex} of the main text, which can be tied up via the $Y_\Psi$ Yukawa coupling. Furthermore we have to include the change in the running of the $\text{SU}(2)_\text{L}$ due to these new states.  Using the methods of Ref.~\cite{Csaki:2023ziz} the estimate for the Majoron mass in Eq.~\eqref{eq:EW} of the main text changes as 
\begin{align}
    m_j^2 \rightarrow m_j^2 \times \left(\frac{M_\Psi}{M_\text{UV}}\right)\times \left(\frac{M_\text{UV}}{M_\Psi}\right)^{\frac{N_S+2}{3}},
\end{align}
where the first term comes from the additional zero modes and the second one from the change of the $\beta$-function. Here $N_S$ denotes the number of additional pairs of scalar doublets with the same mass $M_\Psi$ as the fermions. For $N_S=0$ the additional fermions slightly reduce the estimate for the Majoron mass, and for $N_S=1$ the suppression from the fermionic zero-modes is exactly canceled by the enhancement from the change in the running due to both fermions and scalars as first demonstrated in Refs.~\cite{Nomura:2000yk,Ibe:2018ffn}. The same subtlety might also be relevant for the case of the QCD axion discussed in Ref.~\cite{Csaki:2023ziz}. 

This analysis illustrates that the results and conclusions of the main text that pertain to the estimate of the Majoron mass do not change. If anything the preceding discussion demonstrates that it is more difficult than  naively expected to generate a Majoron mass from electroweak instanton effects, which aligns well with the central message of this work.

\bibliographystyle{utphys}
\bibliography{refs-majoronDW}
\end{document}